\documentclass[
reprint,
aps,pre,notitlepage,nobibnotes,superscriptaddress]{revtex4-2}

\usepackage{color,ifthen,amsthm,thmtools,amsmath,amsxtra,amsfonts,dsfont,graphicx,bm,tikz,scalerel,wasysym,bbm,graphicx,braket,physics,empheq,comment,soul,physics}

\usepackage{subfiles} 
\newif\ifwithSM 
\withSMtrue 

\usepackage[colorlinks=true,linkcolor=blue, citecolor=blue, urlcolor=blue, bookmarks]{hyperref}

\begin{document}
\title{Direct Measurement of the Orderphobic Effect}

\author{O.D. Lunn}
\affiliation{School of Physics and Astronomy, University of Nottingham, Nottingham, NG7 2RD, UK}

\author{J.G. Downs}
\affiliation{School of Physics and Astronomy, University of Nottingham, Nottingham, NG7 2RD, UK}

\author{K.K. Mandadapu}
\affiliation{Department of Chemical and Biomolecular Engineering, University of California, Berkeley, California 94720, USA}

\author{J.P. Garrahan}
\affiliation{School of Physics and Astronomy, University of Nottingham, Nottingham, NG7 2RD, UK}

\author{M.I. Smith}
\thanks{Corresponding author: mike.i.smith@nottingham.ac.uk}
\affiliation{School of Physics and Astronomy, University of Nottingham, Nottingham, NG7 2RD, UK}

\begin{abstract}
	Fluctuation-induced forces, such as the Critical Casimir Effect (CCE) 
	\cite{fisher1978phenomenes, passante2025quantum, gambassi2009the-casimir}, are fundamental mechanisms driving organization and self-assembly near second-order phase transitions \cite{nguyen2013controlling, hoskova2025casimir}. The existence of a comparable, universal force for systems undergoing a first-order transition has remained an unresolved fundamental question. The proposed Orderphobic Effect is one such potential mechanism. It arises from minimisation of the interfacial free energy between solutes that locally nucleate a disordered phase \cite{katira2016pre-transition}. Here, we report the first experimental demonstration and quantitative measurement of the Orderphobic Effect. Using a driven, non-equilibrium quasi-2D granular fluid undergoing a first-order order-disorder transition, we show that specifically designed solutes in an ordered phase nucleate a coexisting ``bubble'' of the disordered phase. By analysing its capillary fluctuations, we confirm that this phenomenon  occurs due to the proximity to phase-coexistence, and we directly quantify the attractive force by measuring the interaction free energy between solutes. The observation of this general fluctuation-mediated force in a non-equilibrium steady state strongly supports its claimed universality. Our work establishes the Orderphobic Effect as the first-order equivalent to the CCE, providing a new, general route for controlling self-assembly and aggregation in soft matter and non-equilibrium systems.
\end{abstract}

\maketitle

\smallskip
\noindent
{\bf Introduction.}
The organisation and collective behaviour of nanoscale materials, from biological membranes to colloidal suspensions, is fundamentally governed by forces arising from thermal and structural fluctuations \cite{israelachvili1982the-hydrophobic, israelachvili1983molecular, kaplan1994entropically, musevic2006two-dimensional, chandler2005interfaces, hertlein2008direct, han2017effective}. These interactions modify processes like self-assembly, pattern formation and phase separation \cite{poon1999colloid-polymer, soyka2008critical, nguyen2013controlling}. An important example is the Critical Casimir Effect (CCE), where the confinement of long wavelength modes near a second-order phase transition induces a force between surfaces \cite{fisher1978phenomenes, nguyen2013controlling}. Experiments played an important part in understanding this profound link between statistical mechanics and nanoscale forces \cite{hertlein2008direct}, revealing that the forces can be tuned through small variations in temperature and surface wettability \cite{soyka2008critical, hertlein2008direct}.

The CCE provides the established paradigm for fluctuation-driven interactions at or near second-order (continuous) phase transitions \cite{fisher1978phenomenes, hertlein2008direct}. However, the existence of an equivalent general force near a {\em first-order} phase transition remains an unresolved question in statistical physics. The hydrophobic effect is a known, {\em specific} example, of a first-order fluctuation-driven interaction. But this applies only to hydrophobic moieties in water \cite{israelachvili1983molecular, chandler2005interfaces, meyer2006recent}. Reference \cite{katira2016pre-transition} proposed an analogous but generic phenomenon, and called it the ``Orderphobic Effect''. This too is a fluctuation-mediated force proposed to arise between solutes that locally nucleate a disordered phase within an otherwise ordered phase; minimisation of the created interface’s free energy drives the solutes together. Since its proposal, the effect has been largely explored through simulations (or in simplified models) \cite{katira2016pre-transition, katira2018solvation, klobas2024exact}, with these results indicating that it is a universal feature of any system close to a first-order order-disorder transition, both static and dynamic. While recent experimental work on hydrophobic mismatch in synthetic lipid membranes is suggestive of this mechanism \cite{peruzzi2024hydrophobic}, the Orderphobic Effect remains  untested. Here we provide the first direct experimental evidence of the Orderphobic Effect, confirming its universal features.

In two-dimensional (2D) hard-sphere-like systems, such as many experimental colloidal \cite{thorneywork2017two-dimensional}, hydrogel \cite{han2008melting}, or granular \cite{reis2006crystallization} systems, the transition from liquid to crystal proceeds via an intermediate hexatic phase \cite{halperin1978theory, nelson1979dislocation-mediated}. This results in the well known two-step defect-mediated melting process corresponding to the binding-unbinding of dislocation and disclination pairs \cite{nelson1979dislocation-mediated, bernard2011two-step}. This often makes the prerequisite one-step, first-order liquid-crystal transition inaccessible in common 2D experimental platforms. However, in Ref.~\cite{downs2021topographic} some of us surmounted this challenge by showing that a quasi-2D vibrated granular fluid could undergo a one-step first-order transition provided the surface was topographically patterned (see Methods), allowing us to directly test the Orderphobic Effect.

Our experiments use the same system as Ref.~\cite{downs2021topographic}: highly inelastic $d=4{\rm mm}$ particles (“the fluid”) vibrated on this patterned surface. We use custom 3D printed orderphobes (“solutes”) designed to be incommensurate with the crystalline particle phase (see Methods and Supplementary Fig.~1). When placed in the crystalline particle layer, these appear to nucleate a local ``bubble'' of the liquid (disordered) phase. Using capillary wave theory, we firstly confirm that the orderphobe is indeed orderphobic, and that the disordered region is a genuine phase-coexistence phenomenon. We then proceed to demonstrate, qualitatively and quantitatively, that there is indeed an attractive orderphobic force, culminating in the direct measurement of the free energy as a function of orderphobe separation. The fact that these novel experiments are also highly non-equilibrium, further strongly supports the claim that the Orderphobic Effect is a truly universal first-order equivalent to the Critical Casimir Effect.

\smallskip
\noindent
{\bf The orderphobe nucleates a disordered phase within the crystal.}
Placing an orderphobe into the 2D granular fluid at phase coexistence, we observe that as it diffuses, it remains invariably associated with a local ``bubble'' of pre-melted liquid. This suggests that the orderphobe favours the disordered phase. Other observations also support this. Firstly, with the vibrations switched off, we carefully remove the minimum number of particles from the pure crystalline phase to insert an orderphobe. As we increase the vibrational amplitude, $A$, particles adjacent to the orderphobe are ejected from the layer and a bubble of disordered fluid forms. This local phase change does not occur anywhere else in the tray or, for equivalent acceleration ($\Gamma = A \omega^2 / g$), in the orderphobe’s absence. This confirms that it is the sole nucleation site.

Secondly, taking a single orderphobe, we loosely constrain its lateral motion using a vertical pin attached to a horizontal glass pole suspended above the shaker. We “heat” the fluid until it is entirely liquid, before slowly “cooling” the system. In all experiments, regardless of the orderphobe’s position, a bubble of disorder forms around it, with the rest of the system being in a uniform crystalline phase. This demonstrates that the liquid (i.e., the disordered phase) preferentially wets the orderphobe.

\begin{figure}[t]
	\includegraphics[width=\columnwidth]{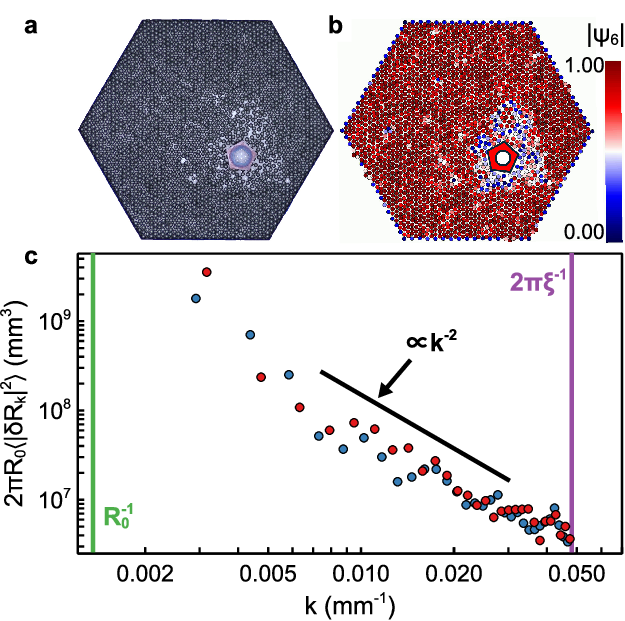}
	\caption{
		{\bf Experimental results: capillary fluctuations.} a) Example image of an orderphobe in the particle monolayer. b) superimposed local orientational order parameter, $|\psi_6^i|$. c) Fourier components of the interfacial radial fluctuations $\delta R_k$ for an orderphobe-nucleated bubble (blue) and one formed simply due to phase-coexistence (red) of the same size. The green vertical line indicates $k=2\pi/(2\pi R_0)$, where $R_0$ is the average radius of the interface. The purple vertical line corresponds to the coarse graining width, $k=2\pi/\xi$. The black line indicates $k^{-2}$ scaling.
		}
\end{figure}

\smallskip
\noindent
{\bf Capillary fluctuation measurements.} To demonstrate unequivocally that the bubble around the orderphobe forms due to nucleation of a coexisting phase, we study its interfacial capillary fluctuations. This enables us to confirm the underlying mechanism is not due to, for example, geometric frustration or orderphobe inertia, which may be relevant for granular systems.

We first prepare the system at coexistence without the orderphobe, where under steady state conditions the crystal and liquid coexist. We observe a single small bubble of disorder (see Fig.~ 1a). The interface separating these two phases undergoes radial fluctuations. To analyse this, we track the position of every particle and calculate the local hexatic order parameters $|\psi_6^i|$ (see Methods). We then use the order parameter to identify the interface between the ordered and disordered phases (see Fig.~1b). Each point along the interface fluctuates by an amount $\delta R$ relative to the time averaged radius $R_0$. Taking the Fourier transform of the interfacial profile, we characterise these capillary fluctuations as a function of wave number $k$ (see Fig.~1c). 

Equilibrium capillary wave theory (CWT) predicts interfacial fluctuations which scale proportional to $k^{-2}$, and whose magnitude depends on the interfacial tension $\gamma$ and temperature $T$ \cite{fisher1982agreement, aarts2004direct}. CWT relies on the system being in equilibrium, which is manifestly not the case under the non-equilibrium conditions in this vibrated granular experiment. However, recent theoretical work on active matter has shown that many canonical equilibrium results, including the $k^{-2}$ scaling of capillary fluctuations, can be recovered using an effective temperature ($T_{\rm eff}$), despite the microscopic differences in energy flux \cite{han2017effective, langford2024theory} (see Supplementary Information for a detailed discussion). Fig.~1c shows a rare experimental demonstration that a driven, non-equilibrium system can still obey equilibrium-like capillary scaling \cite{luu2013capillarylike}, thereby confirming the usefulness of an effective temperature in describing this class of systems.

Figure 1c compares the capillary fluctuations of two distinct interfaces: a bubble nucleated by an orderphobe (blue) and another arising from simple liquid-crystal coexistence without any solute (red). The fact that both sets of data fall onto a single curve confirms that the bubble created by the orderphobe is inherently linked to the underlying first-order phase transition between the ordered and disordered bulk phases. We thus conclude that the solute is orderphobic: it nucleates a bubble of the liquid phase due to thermal-like fluctuations close to a first-order phase transition.

\begin{figure}[t]
	\includegraphics[width=\columnwidth]{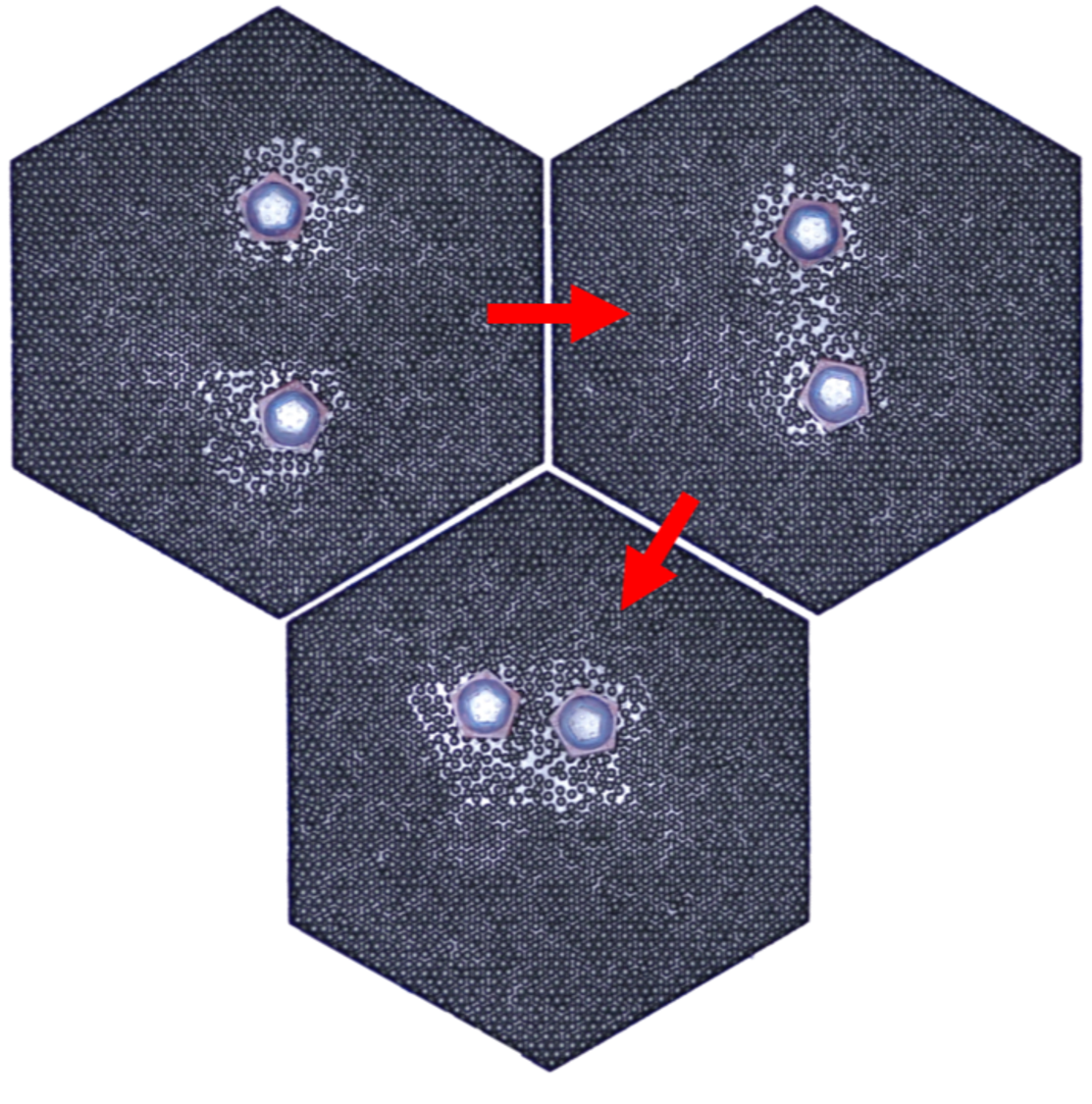}
	\caption{
		{\bf Qualitative demonstration of the Orderphobic Effect.} Experimental images of two solutes in an otherwise ordered system. Red arrows indicate progression of time.
		}
\end{figure}

\smallskip
\noindent
{\bf 
Experimental demonstration of the Orderphobic Effect.} 
As the system is slowly cooled through its first-order crystal-liquid phase transition, a small bubble of disorder is maintained in the vicinity of the two solutes.  Minimisation of the interfacial energy associated with the interface makes two isolated bubbles inherently unstable. Reference~\cite{katira2016pre-transition} showed through simulations of proteins embedded in lipid membranes that as the bubbles of disorder diffuse into proximity, fluctuations result in the interfaces merging: initially an elongated capillary bridge forms; subsequently, capillary pressure rounds the merged interface. This generates an effective attractive force that draws the two solutes together to reduce the net interfacial energy, a predicted hallmark of the Orderphobic Effect. 

To study the Orderphobic Effect experimentally, we initially position two orderphobes approximately 80mm apart. The translational motion of the orderphobes is constrained by vertical pins attached to a glass rod above the tray. We slowly cool the granular fluid into the ordered phase ($\Gamma \sim 2.71 \pm 0.02$). Under these conditions each orderphobe is surrounded by a small distinct bubble of disorder. The glass rod is then removed allowing the orderphobes to diffuse (see Fig.~2 and Supplementary Video 1).
 
There are three possible experimental outcomes: (i) orderphobes sometimes ``pop out'' of the particles layer terminating the experiment, (ii) the bubbles diffuse apart towards the boundary, (iii) the bubbles encounter one another. When the bubbles first meet, the fluctuating interfaces form a narrow strip of disorder which may intermittently break and reform. However, once a significant bridge is established, the bubbles never separate (based on 17 repeat experiments). We find that the rounding of the interface is significantly slower than the initial bridge formation. Despite the vast differences in length and timescales—and our system being far from equilibrium—these results replicate the qualitative features of the Orderphobic Effect previously observed in simulations of membrane proteins \cite{katira2016pre-transition}.

Figure 3a shows a schematic of an experimental setup, which mimics for our system an approach often used with optical tweezers, where fluctuations in the position of a colloidal particle, constrained by a potential, can be used to quantify an underlying force \cite{crocker1996microscopic}. As shown in Figures 3a and 3b, two vertical pins on a horizontal glass rod constrain the orderphobes via a circular recess in their Perspex tops, limiting their travel. Tracking each orderphobe we measure their separation, $x_{\rm i}$, relative to the separation of the pins ($x_{\rm p}$ = 60mm): from change in the distribution of separations we can directly infer the force between the solutes. To establish a baseline and account for the system's slow relaxation, we individually tracked each orderphobe on its respective pin for 30 hours ($\Gamma \sim 2.61, \phi=0.84$). These control experiments yielded symmetric probability distributions, confirming the absence of systematic bias in the setup.

The Orderphobic Effect arises from the minimisation of the interfacial free energy of a single bubble of disorder containing two orderphobes. We therefore performed a critical control experiment by reducing the number of particles in the experiment ($\phi=0.69$), thereby shifting the effective temperature at which the liquid-crystal transition occurs. Shaking the experiment, at the same value of $\Gamma \sim 2.61$ but with no liquid-crystal interface present, we measure the separation between the two orderphobes. Fig.~3c (blue line) shows that the distribution is completely symmetric indicating that there are no measurable interactions when the system is entirely in one phase.

Restoring the particle concentration to $\phi=0.84$ while maintaining $\Gamma \sim 2.61$ results in the formation of a single, elongated bubble containing both orderphobes. After allowing 20 minutes for the system to reach steady-state, we then collected images for 54 hours. Fig.~3c (red line) shows the measured probability distribution of the relative separation. The curve exhibits a pronounced bias towards smaller separations, indicating the presence of an attractive force between the orderphobes. That this force is only observed when a disordered bubble is nucleated in the crystal phase confirms that the force is directly related to the underlying first-order phase transition. We repeated each experiment, and in each case the bias was only seen with the merged bubble. This provides quantitative experimental evidence of the orderphobic force. Furthermore, the observation of this force in a non-equilibrium steady state also strongly supports the assertion that this is a universal phenomenon for solutes in systems that undergo a first-order order-disorder phase transition.

\begin{figure}[t]
	\includegraphics[width=\columnwidth]{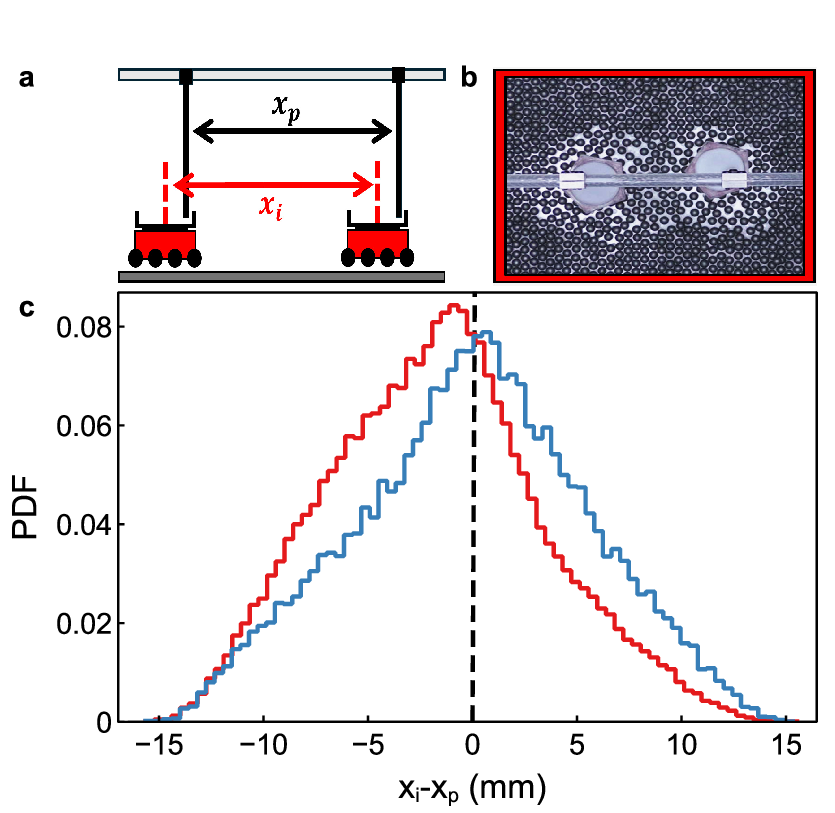}
	\caption{
		{\bf Quantifying the effect of an orderphobic force.} a) Schematic demonstrating the experimental setup. b) Image of the experiment from above. c) Probability distribution of orderphobe separations with (red) and without (blue) an order-disorder interface. The interface nucleated by the orderphobes results in a bias in the histogram towards smaller separations, demonstrating the existence of an attractive orderphobic force.
		}
\end{figure}

\smallskip
\noindent
{\bf 
Measuring how the free energy changes with separation.}
While the distribution bias in Fig.~3c provides a quantitative demonstration of the Orderphobic Effect, a more rigorous approach is to extract the full free energy profile as a function of orderphobe separation, $D$. This profile is governed by the free energy of the liquid-crystal interfaces surrounding the solutes. To that end, we constrained two orderphobes at fixed separations ranging from 35–125 mm. While their positions were fixed by the pins, they remained free to rotate. For each separation $D$, we measured the fluctuating interface length, $L$, over time. Fig.~4a shows representative histograms of these fluctuations. At small separations, the system consistently forms a single bridge of disorder (one bubble), which minimises the free energy. However, as $D$ increases, the free energy minimum shifts, causing a transition into two distinct bubbles. In the intermediate regime, the system fluctuates intermittently between the one-bubble and two-bubble states (see Fig.~4c).

The fixed separations, $D$, artificially constrain the system, producing the narrow, localized distributions of interfacial length $L$ shown in Fig.~4a. To reconstruct the free energy profile for freely diffusing orderphobes, we reweight these distributions using the Multistate Bennett Acceptance Ratio (MBAR) approach \cite{shirts2008statistically}, see Methods. This umbrella sampling technique uses the overlap between intersecting histograms to determine the free energy differences between neighbouring states, yielding an estimate of the  single, globally unbiased, distribution of interface lengths (see Fig.~4b). 

From this systematic analysis, we extract the total free energy of interaction, $\Delta F / k_{\rm B} T_{\rm eff}$ (see Fig.~4c, red circles): it grows approximately linearly with separation $D$ until a critical separation, whereupon it plateaus. The range of this interaction is particularly instructive, confirming that the Orderphobic Effect is fundamentally a fluctuation-driven phenomenon.

By measuring the equilibrium radius $R_{0}$ and the standard deviation of radial fluctuations $\delta R$ for isolated bubbles, we find that the range of the force extends significantly beyond the static contact distance $D = 2 R_{0}$. Specifically, the interaction persists until the separation reaches approximately $2(R_{0} + 2 \delta R)$, as indicated by the hashed regions in Fig.~4c. This suggests that the attractive force effectively vanishes only when the gap between the bubbles exceeds the distance that can be bridged by a simultaneous two-standard-deviation fluctuation in each bubble's radius.

Analysing all the images at each separation, we  measure the total amount of time $t_1$ the system spends in a one-bubble configuration, and the time $t_2$ for a two-bubble configuration (see Fig.~4c, blue circles). The change in the gradient in free energy occurs for the same values of $D$ (see shaded region in Fig.~4c) as the rapid change of $t_{1}$  indicating bubble separation. This conclusively demonstrates that the measured force is tied to interactions between the fluctuating interfaces nucleated by each orderphobe for the system in the vicinity of the first-order transition.

\begin{figure*}[t]
	\includegraphics[width=\textwidth]{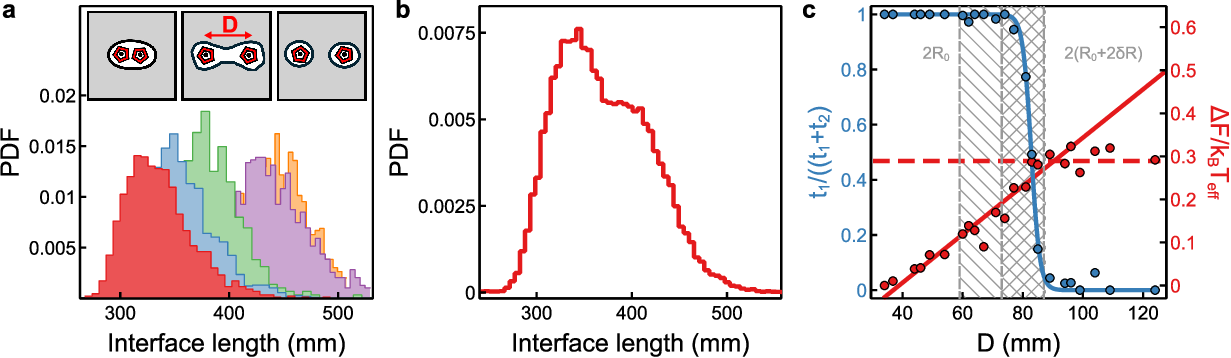}
	\caption{
		{\bf Measuring the Free energy change due to the Orderphobic Effect.} 
		a) Subset of the distributions of the interfacial length for increasing pole separations, $D$. Inset, schematic illustrating the transition from one to two bubbles as the pin separation, $D$, is increased. 
		b) Distributions of interface lengths (red line), from MBAR reweighting (see Methods).
		c) Proportion of time spent as one ($t_1$) or two bubbles ($t_2$) (blue circles, left scale) and the scaled free energy, $\Delta F / k_{\rm B} T_{\rm eff}$ (red circles, right scale) as a function of pole separation, $D$. Solid / dashed red lines are guides for the eye. The grey single hashed and double hashed regions are bounded by $2 R_{0}$, $2(R_{0}+\delta R_{0})$ and $2(R_{0}+2 \delta R)$ representing the distances at which the bubbles can meet via one and two standard deviations of their radial fluctuations respectively.
		}
\end{figure*}

\smallskip
\noindent
{\bf 
Discussion.}
The Critical Casimir Effect is the paradigmatic fluctuation induced force associated with second-order phase transitions \cite{hertlein2008direct, gambassi2009the-casimir, passante2025quantum}. We have systematically demonstrated that the Orderphobic Effect, first proposed theoretically in Ref.~\cite{katira2016pre-transition}, is an equally general force associated with vicinity of first-order phase transitions. In our experiments we have confirmed that an appropriate solute which prefers the disordered phase can, when placed in a homogeneous crystal phase, nucleate a coexisting bubble of the disordered phase. We further showed that the interaction between the interfaces around two such solutes produces a fluctuation induced attractive force driven by minimisation of the interfacial free energy.

The parallels between the Orderphobic Effect and CCE extend to our understanding of how such forces may be controlled. Both effects, arising from their proximity to a transition, are sensitive to temperature \cite{hertlein2008direct}. In the CCE, the modification of surface wetting properties can be used to tune, or even invert the sign of, the fluctuation induced forces \cite{hertlein2008direct, schmidt2022tunable, nguyen2013controlling}. Similarly, in Ref.~\cite{downs2021topographic} we have shown that modifying the microstructure of boundaries can invert the wetting behaviour producing a preference for the ordered phase. This suggests that in the Orderphobic Effect modifying surface wetting properties, as for the CCE, could be used to tune or even invert the strength of the effect, a concept tested in simulations, but not yet realised experimentally. 

Finally, the general nature of this interaction suggests its relevance far beyond conventional soft materials. Our observation in a non-equilibrium steady state system supports the idea that it is sufficiently general. It would be intriguing to explore the Orderphobic Effect near first-order phase transitions in other systems where such first-order transitions are known to exist. Our results here establish the Orderphobic Effect as a general principle, opening a new avenue for exploiting fluctuation-driven interactions in materials science and statistical physics.

\bigskip
\noindent
{\bf 
METHODS.} Our experiments use the quasi- 2D vibrated granular system described in Ref.~\cite{downs2021topographic}, which enables the study of first-order phase transitions. The fluid consists of Nitrile particles ($d=4 {\rm mm}$), with a low coefficient of restitution ($\varepsilon \sim 0.1$), placed on a horizontal aluminium plate subject to small vertical vibrations. The plate surface is patterned with a triangular array of shallow dimples (0.16mm deep, 4.8mm spacing), which induces a one-step first-order liquid-crystal phase transition as the vibration amplitude is decreased \cite{downs2021topographic}. Particles, attached to the 3D printed boundaries were used to control the wetting behaviour of the granular fluid, ensuring the ordered phase forms adjacent to the boundary. 

Orderphobes (solutes) were 3D printed from PLA in a pentagonal shape. Along each side, particles were partially recessed, with a spacing chosen to maximise their incommensurability with the crystalline phase. To damp the orderphobe’s vertical motion a transparent Perspex lid of comparable mass was loosely placed on the orderphobe body.
The system was filmed from above with a camera (Panasonic HCX1000, 4k, 50fps). Particle positions were tracked using the OpenCV Hough circle transform.

\smallskip
\noindent
{\bf Capillary fluctuations analysis.}
The hexatic order parameter $\psi_6^i = \frac{1}{N_j} \sum_{j=1}^{N_j} e^{6i \theta_{ij}}$, is calculated for each particle using its coordinates, and the number of nearest neighbours ($N_j$) determined by Delauney triangulation. To identify the interface between phases the local order parameters are coarse grained following the method described in Ref.~\cite{katira2016pre-transition} using a convolution with a normalised Gaussian kernel $G(\vec{r}-\vec{r}_i)$ and coarse-graining width $\xi$
\begin{equation}
	\overline{\psi}(\vec{r}) = \sum_i G(\vec{r}-\vec{r}_i) |\psi_6^i| ,
\end{equation}
The coarse-graining width, $\xi$ is chosen to be 1.5a where a is the spacing of particles in the ordered phase. Values of $\xi$ in the range, $1.5a<\xi<3a$ accurately pick out the interface. The threshold for the interface is defined as 
$\overline{\psi}(\vec{r}) = (\overline{\psi}_{\rm O} + \overline{\psi}_{\rm D})/2$
where $\overline{\psi}_{\rm O}$ and $\overline{\psi}_{\rm D}$ are the averages of the coarse-grained field in each phase.

\smallskip
\noindent
{\bf MBAR analysis.} At each separation $D$, we collected $\sim$~1000 images of the interface at intervals of 10s. The same interface finding procedure was employed as above and the length of the contour extracted using OpenCV. These samples of the interface length were collected at 23 values of $D$ ranging from 35-125mm. The MBAR algorithm \cite{shirts2008statistically} was then used to calculate the free energy differences between these states, minimising the statistical error across the entire dataset.

\bigskip
\noindent
{\bf Acknowledgements.} O.D.L.\ acknowledges an EPSRC Studentship (EP/W524402/1). Funding from the Leverhulme Trust is gratefully acknowledged (RPG-2025-017). We thank J. Dorman and M.R. Swift for insightful discussions.

\smallskip
\noindent
{\bf Contribution Statement.}
M.I.S., K.K.M., J.P.G.\ conceived the project. M.I.S.\ designed the experiments. J.G.D.\ performed preparatory experiments. M.I.S., J.G.D., O.D.L.\ developed the experimental methodology. O.D.L.\ performed the main experiments. O.D.L.\, M.I.S.\ performed the analysis. All authors contributed to discussion, interpretation of the results, and writing of the paper.

\smallskip
\noindent
{\bf Competing Interests.} The authors declare no competing interests.

\smallskip
\noindent
{\bf Data availability.}
The particle tracking data in all the experiments included in this study, together with explanations of the datasets and python code used for analysis is provided at \href{https://doi.org/10.5281/zenodo.18153363}{DOI:10.5281/zenodo.18153363}.

\bibliographystyle{apsrev4-2}

\bibliography{references}

\ifwithSM
  \clearpage

\title{Direct Measurement of the Orderphobic Effect}

\author{O.D. Lunn}
\affiliation{School of Physics and Astronomy, University of Nottingham, Nottingham, NG7 2RD, UK}

\author{J.G. Downs}
\affiliation{School of Physics and Astronomy, University of Nottingham, Nottingham, NG7 2RD, UK}

\author{K.K. Mandadapu}
\affiliation{Department of Chemical and Biomolecular Engineering, University of California, Berkeley, California 94720, USA}

\author{J.P. Garrahan}
\affiliation{School of Physics and Astronomy, University of Nottingham, Nottingham, NG7 2RD, UK}

\author{M.I. Smith}
\thanks{Corresponding author: mike.i.smith@nottingham.ac.uk}
\affiliation{School of Physics and Astronomy, University of Nottingham, Nottingham, NG7 2RD, UK}

\maketitle

\onecolumngrid

\begin{center}
    {\large \bf SUPPLEMENTARY MATERIAL}
\end{center}

\section{Supplementary Video 1}

Supplementary video 1 shows two orderphobes coming together under the influence of the attractive Orderphobic Effect. The orderphobes are initially constrained via vertical pins attached to a horizontal glass rod, 80mm apart. The system is allowed to equilibrate for 20mins after which time the rod is removed. The orderphobes diffuse throughout the experiment. As their fluctuating interfaces merge, a single bubble is formed. This is initially elongated but gradually the weak interfacial tension rounds the bubble. This provides an effective attractive interaction between the orderphobes which pulls them together. The speed of the movie is increased by 16x. The entire movie lasts 960s in real time.

\section{Methods}

\begin{figure}[h]
    \centering
    \includegraphics[width=0.5\linewidth]{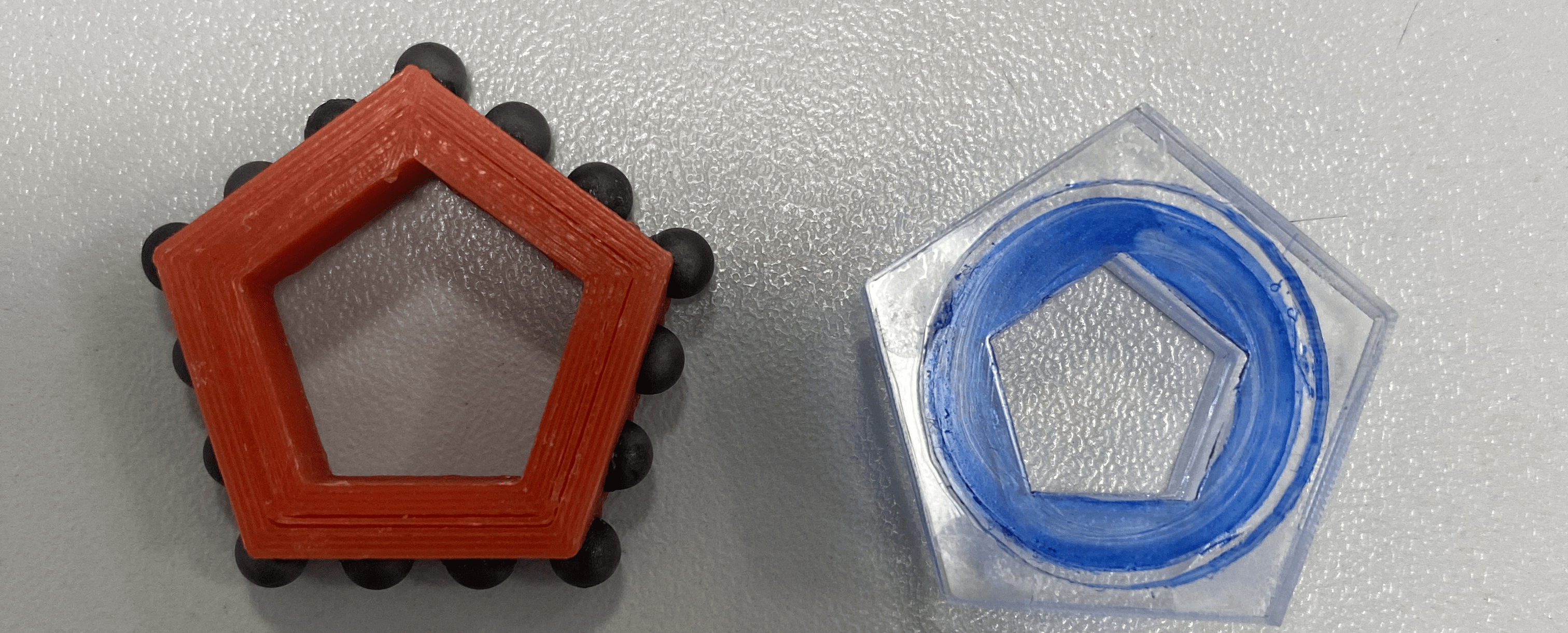}
    \caption{
        Image showing the orderphobe (left) and its lid (right). The orderphobes are 3D printed PLA pentagons with nitrile particles attached to the edges of the pentagon. The lids are laser cut from perspex.
        }
    \label{fig:intruder_hat}
\end{figure}

\begin{figure}[h]
    \centering
    \includegraphics[width=0.45\linewidth]{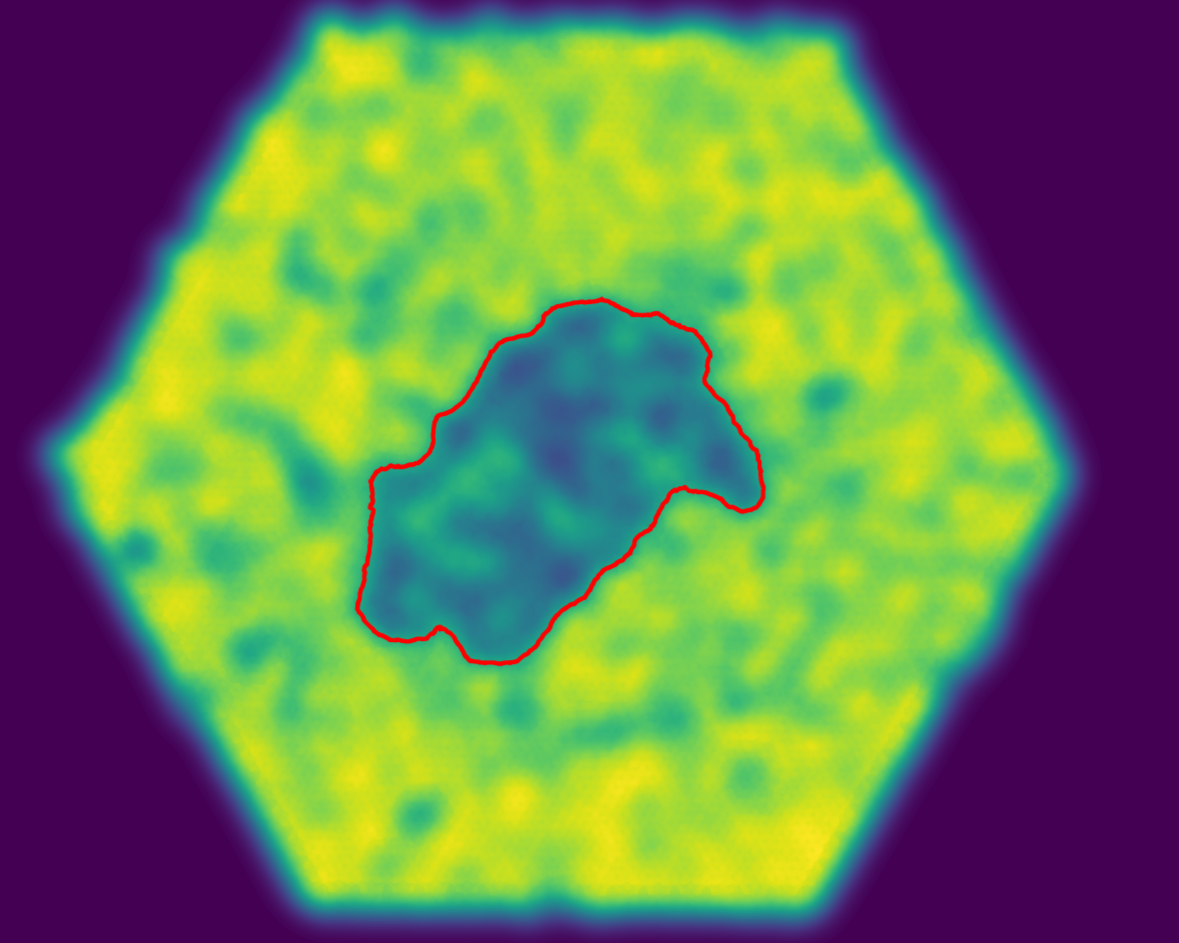}
    \caption{
        An example of the coarse-graining procedure used to find the order-disorder interface. Figure shows the coarse-grained field, $\bar\psi(\vec{r})$. The red line corresponds to the interface detected at $(\psi_{o}+\psi_{d})/2$.
        }
    \label{fig:interface}
\end{figure}

\subsection{Interface detection}

The instantaneous order-disorder interface is found at points where $\bar{\psi}(\vec{r}) = (\psi_{o}+\psi_{d})/2$. $\psi_{o}$ and $\psi_{d}$ are the averages of the magnitude of the local orientational order parameter in the ordered and disordered phases respectively. An example of the coarse-grained field, $\bar{\psi}(\vec{r})$ is shown Supplementary Fig.~\ref{fig:interface}. The interface between the ordered and disordered phases is denoted by the red line.

\subsection{Capillary fluctuations}

Recent work on active matter has shown that many equilibrium results can be recovered provided one chooses an appropriate effective temperature, $T_{\rm eff}$. Reference~\cite{langford2024theory} derive an expression for capillary scaling of active systems. Equation (7) in Ref.~\cite{langford2024theory} consists of two terms: an isotropic (${\sim}k^{-2}$) and an anisotropic component (${\sim}k^{-1}$). Whilst driven granular systems are known to display similarities with active matter, the persistence of motion, whilst present in a limited way \cite{smith2021collision-enhanced} is small. A particle in our liquid phase has a persistence of motion ($\leq{d}$) which is small compared to the wavelength of fluctuations, hence we recover the same capillary scaling as in equilibrium.

\subsection{Quantitative demonstration of the Orderphobic force}

\begin{figure}[h]
    \centering
    \includegraphics[width=0.4\linewidth]{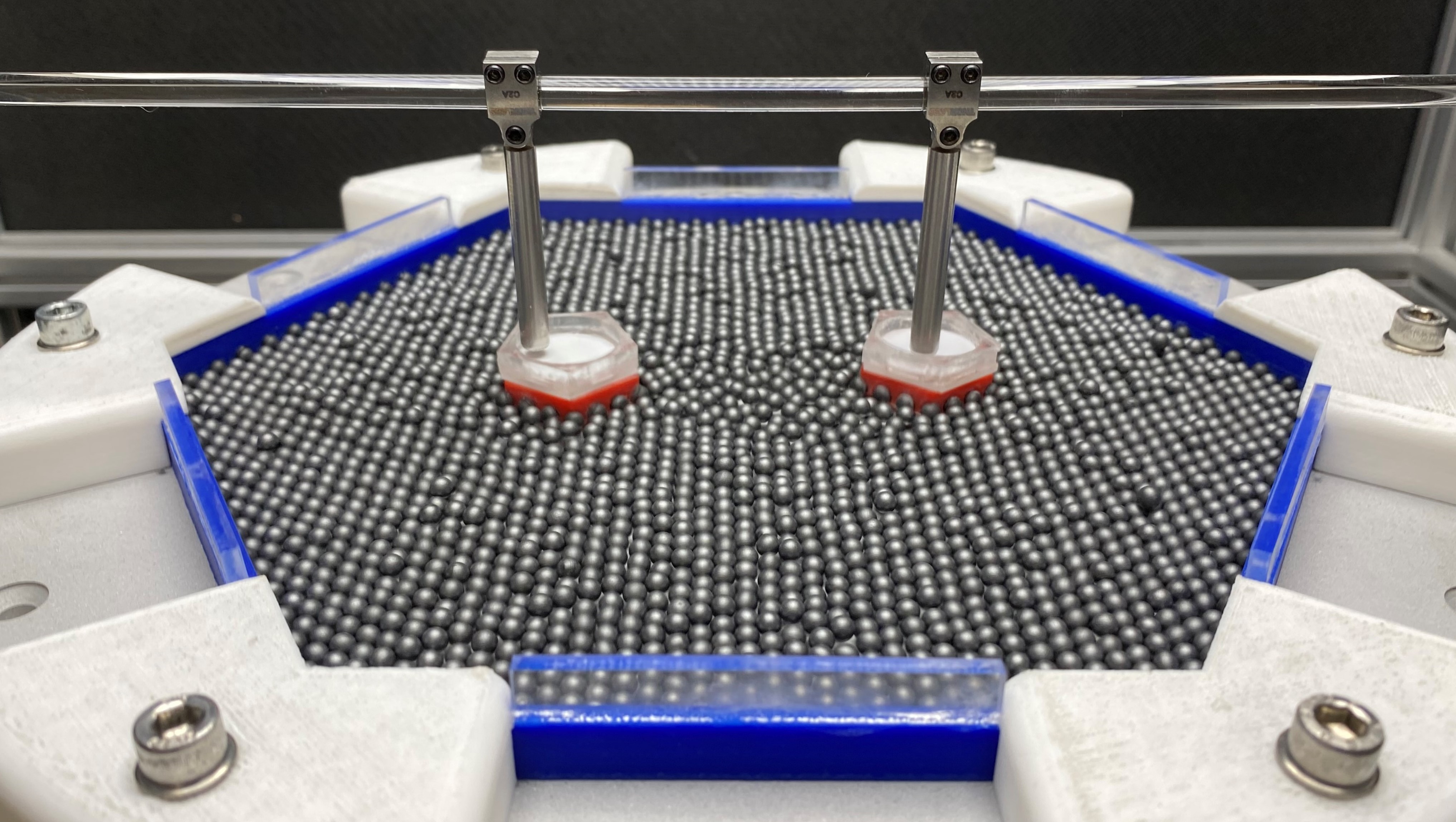}
    \hfill
    \includegraphics[width=0.4\linewidth]{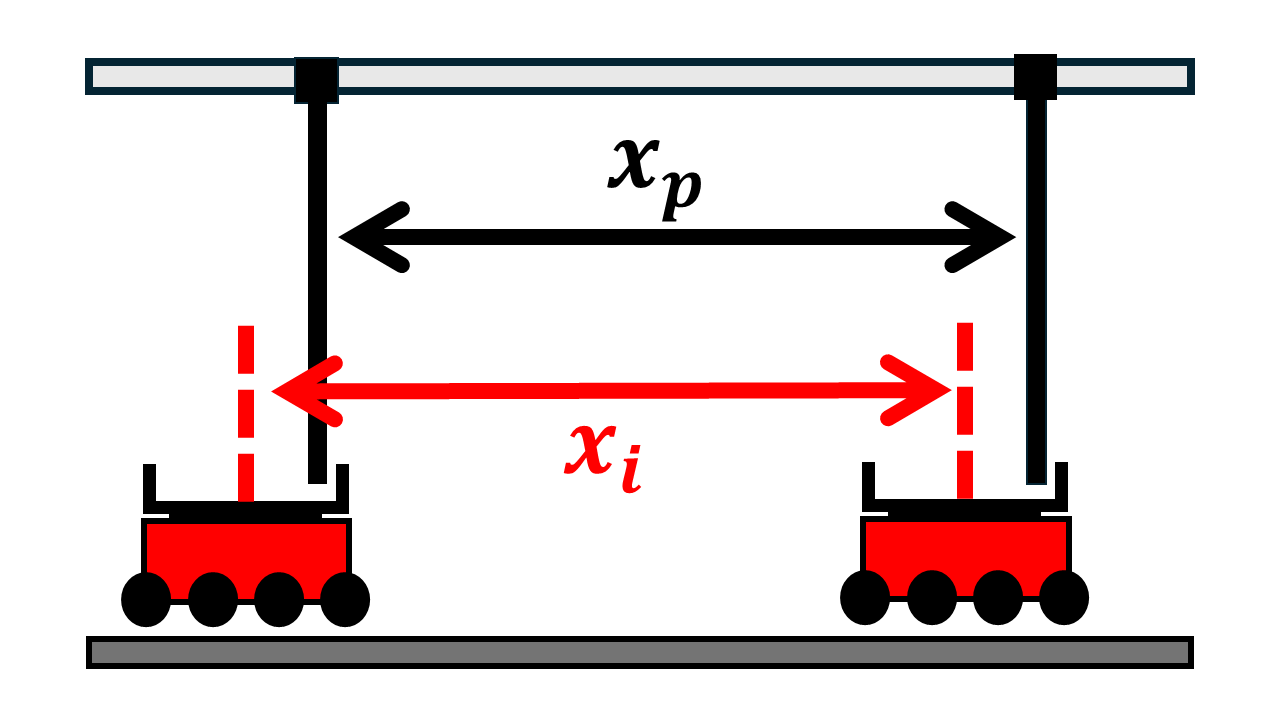}
    \caption{
        (a) Image of the experimental setup, with the two orderphobes loosely confined by pins. 
        (b) A schematic showing the experimental set-up. $x_{p}$ is the distance between the two pins suspended from the glass rod and $x_{i}$ is the distance between the centres of the two orderphobes.
        }
        \label{fig:pole_schematic}
\end{figure}

\subsection{Extracting the free energy from interface fluctuations}

The two intruders are constrained by two pins ($D = 35-125{\rm mm}$) stuck to the surface of the shaker cell. 
The intruders may rotate about the constraining pole, however all translational motion is restricted.
For each pole separation, $D$, we allow the system 20 minutes to reach the steady state before imaging the system every 10s for $\geq$3 hours.

\begin{figure}[h]
    \centering
    \includegraphics[width=0.30\linewidth]{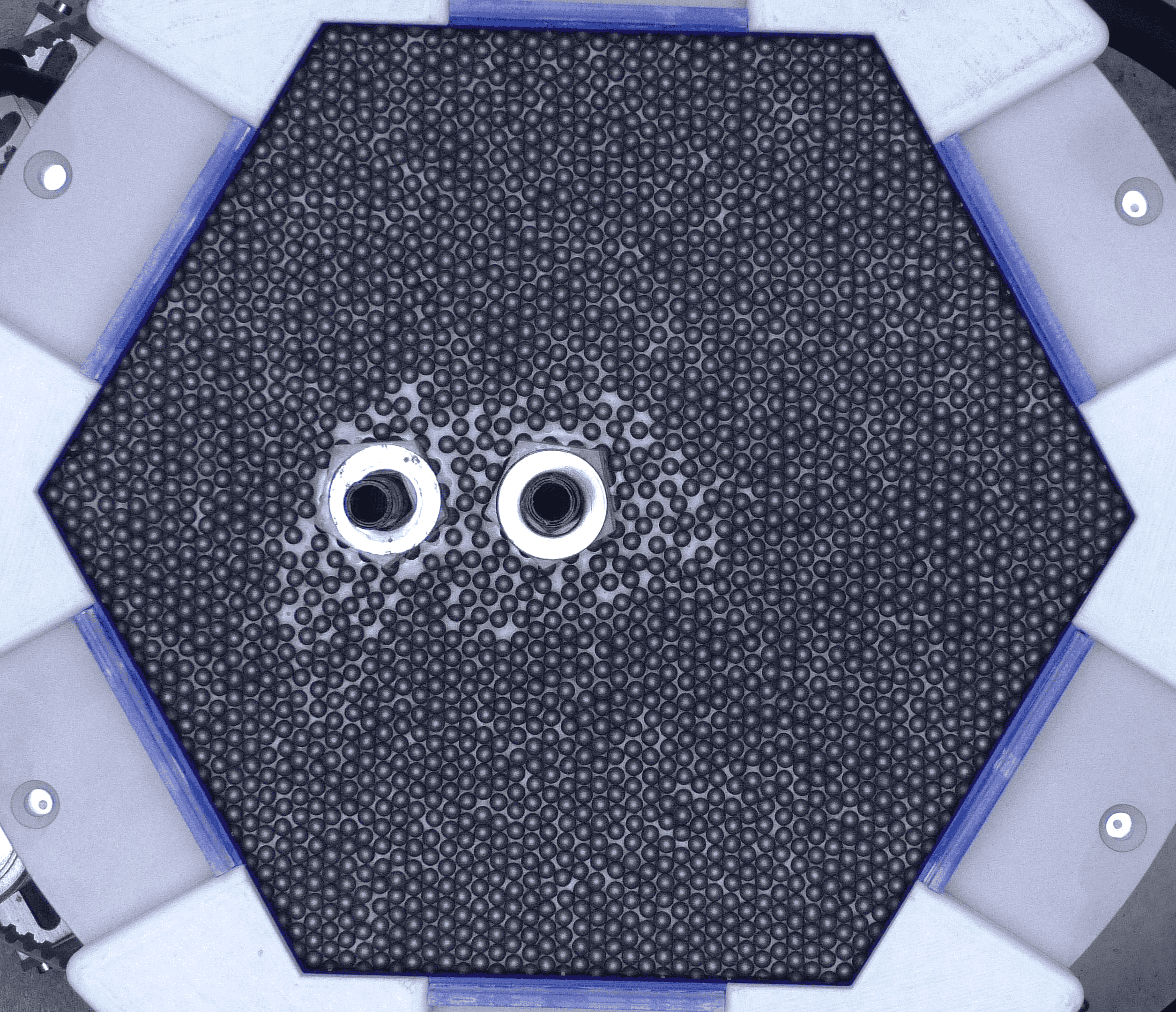}
    \hfill
    \includegraphics[width=0.30\linewidth]{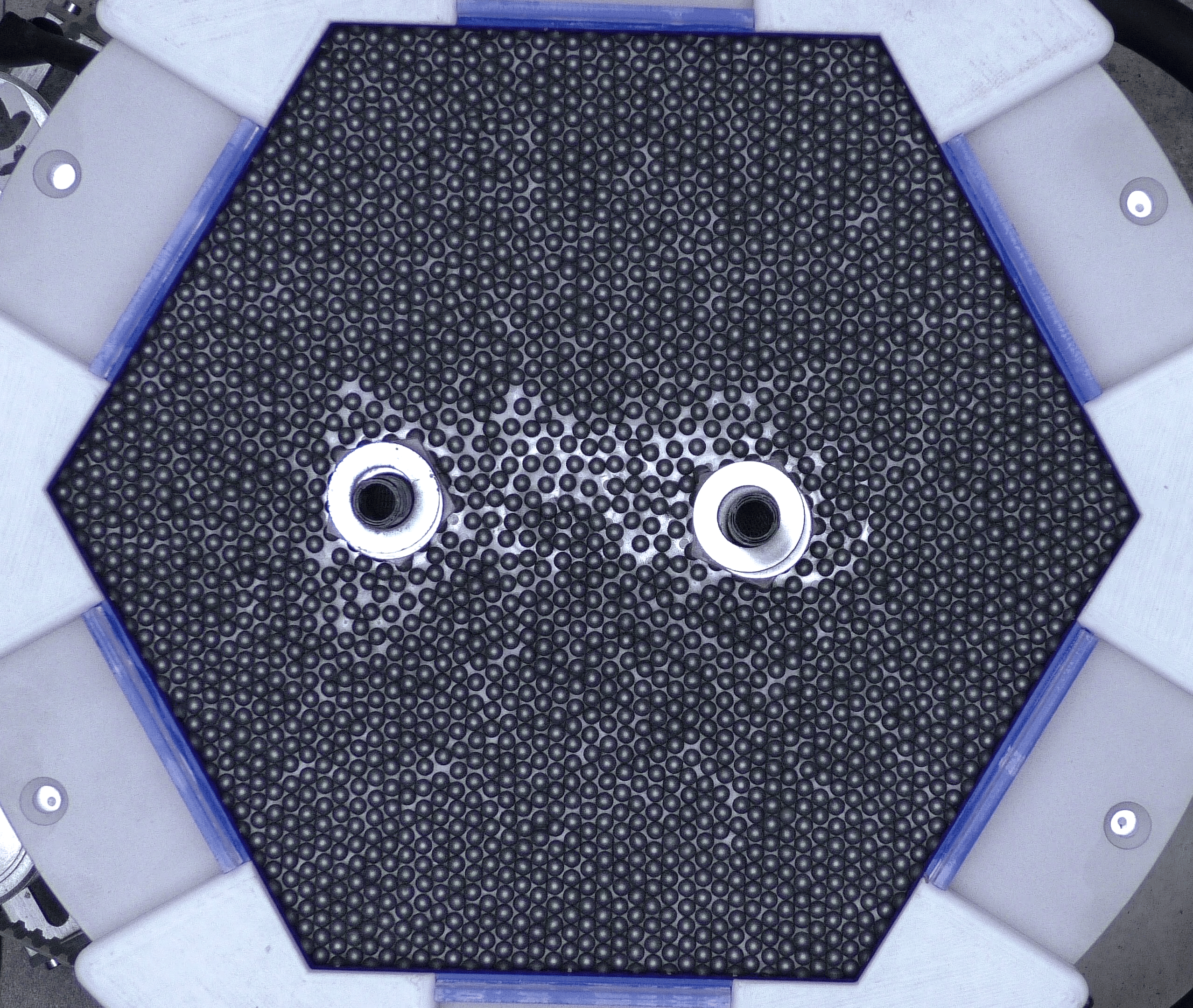}
    \hfill
    \includegraphics[width=0.32\linewidth]{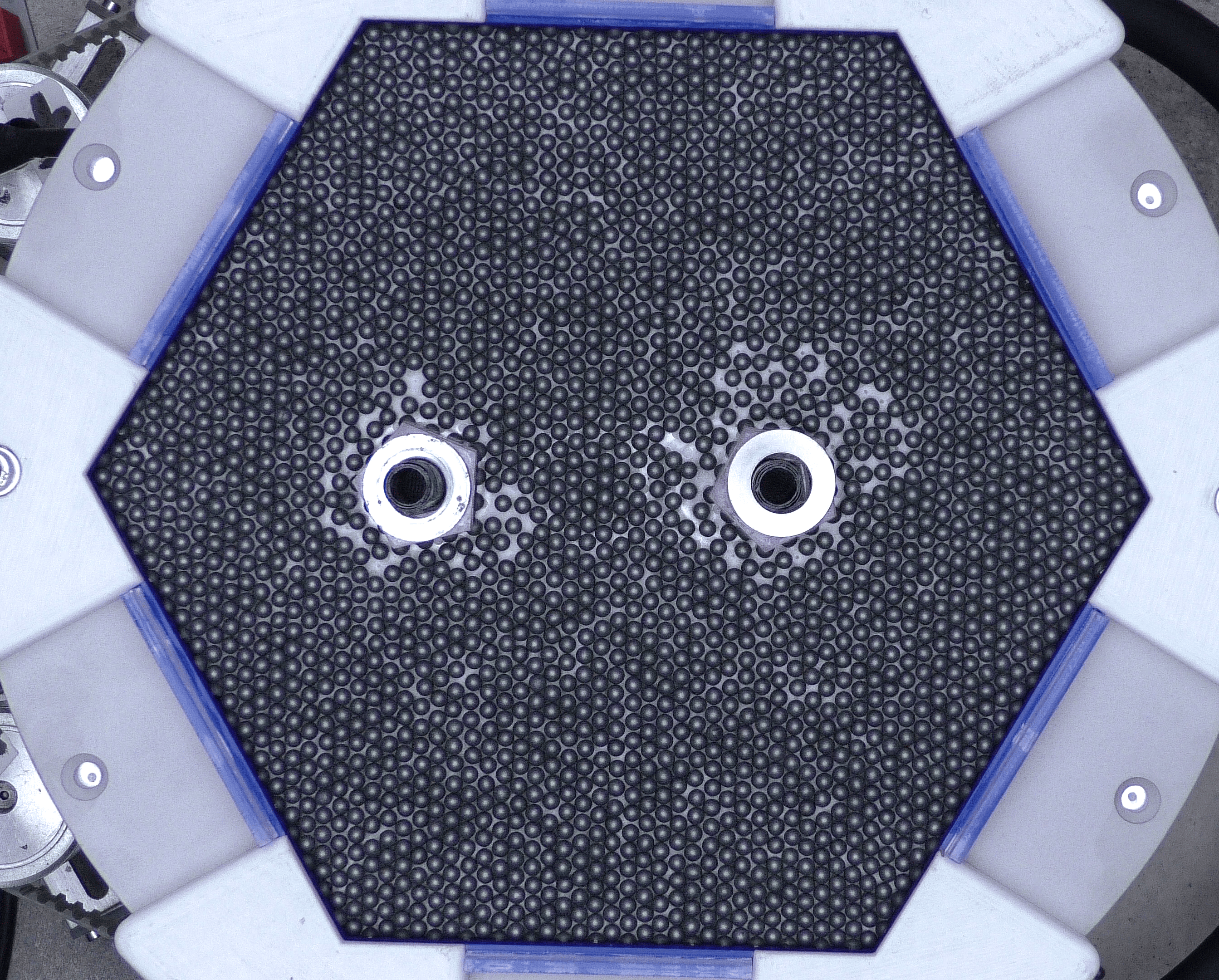}
    \caption{
        Experiments into the behaviour of the interface between two intruders. The intruders are held in place by poles fixed to the surface of the tray, the separation between the poles is changed between experiments.
    }
    \label{fig:int_length_exp}
\end{figure}

\ifSubfilesClassLoaded{
    \bibliographystyle{naturemag}
    \bibliography{references.bib}
}{}

\end{document}

\fi

\end{document}
\typeout{get arXiv to do 4 passes: Label(s) may have changed. Rerun}